\documentclass[aps,prb,showpacs,superscriptaddress,twocolumn]{revtex4}

\usepackage{graphicx}
\usepackage{color}

\begin{document}


\title{Signature of persistent metallic domains in FORC measurements of the VO$_2$ metal-insulator transition}



\author{J. -G. Ram\'irez}
\affiliation{Physics Department, University of California-San Diego, La Jolla California 92093-0319, USA}
\affiliation{Thin Film Group, Universidad del Valle A.A.25360, Cali, Colombia}

\author{A. Sharoni}
\email[]{asharoni@physics.ucsd.edu}
\affiliation{Physics Department, University of California-San Diego, La Jolla California 92093-0319, USA}

\author{Y. Dubi}
\affiliation{Physics Department, University of California-San Diego, La Jolla California 92093-0319, USA}

\author{M. E. G\'omez}
\affiliation{Thin Film Group, Universidad del Valle A.A.25360, Cali, Colombia}

\author{Ivan K. Schuller}
\email[]{ischuller@ucsd.edu}
\affiliation{Physics Department, University of California-San Diego, La Jolla California 92093-0319, USA}

\date{\today}

\begin{abstract}
We have performed first order reversal curve measurements of the temperature-driven metal-insulator transition in VO$_2$ thin films, which enable quantitative analysis of the hysteresis behavior. An unexpected tail-like feature in the contour plot of the reversal curve distribution indicates the existence of metallic domains, even at temperatures below the closing of the hysteresis. These domains interact with the surrounding medium and change the reversal path relative to a path from a \emph{fully} insulating state.  With this in mind, and assuming that such interaction persist through the entire phase transition, we develop a model where the driving force (or energy barrier) in charge of opening a hysteresis in VO$_2$ are inter-domain interactions.  This model is intrinsically different from the Preisach model usually used to describe hysteresis; given that it looks for the microscopic origin of the hysteresis, and provides physical parameters to characterize it.
\end{abstract}


\maketitle


\section{Introduction}

Vanadium oxide (VO$_2$) exhibits a sharp  first-order metal-insulator transition (MIT) and structural-phase transition (SPT) from a high temperature metallic rutile phase to a low temperature insulating monoclinic phase\cite{Masatoshi}, with a hysteresis of $\sim$1K in bulk samples\cite{Duchene:TSF1972,Kim:APL1994} which can reach a few degrees in thin films \cite{sharoni:026404,BrassardAPL2005}.

	The nature of the electronic and structural phase transition has been a matter of research and debate for over three decades, with many recent experiments shedding additional light on the issue \cite{Cavalleri:PRL2001,Cavalleri:PRL2005,Cavalleri:PRB20041,Cavalleri:PRB2004, Qazilbash:PRB2008,Qazilbash:APL2008,Qazilbash:science2007, Driscoll:APL2008,Kim:APL2007,Lee:APL2008}. But during this period the hysteretic nature of VO$_2$ has received much less attention.
		
	The hysteresis nature and mechanism in VO$_2$ is still an open question which is interesting from a basic research view point, and might provide both information regarding the MIT mechanism in VO$_2$ and of hysteresis in general. It is also of vast importance from a technological aspect. For example, for optical memory-type applications, a large hysteresis is needed, while a small hysteresis is preferable in VO$_2$ based room temperature bolometers for UV detectors. So, understanding the origin of the hysteresis can lead to a better control of its properties.
	The VO$_2$ thermal hysteresis characteristics are found to vary between samples grown under different conditions, such as thickness of thin films\cite{Suh:JAP2004}, growth temperature\cite{Youn:JAP:2004}, grain size\cite{Brassard:APL2005}, and choice of substrate \cite{Garry:TSF2004,Kim:APL1994}. For example, in polycrystalline thin films and single crystals of VO$_2$ embedded in silicon, the hysteresis width grows with decreasing of the crystallite size\cite{Brassard:APL2005}.
	Attempts in modeling the thermal hysteresis have been mainly by various forms of the Preisach model\cite{Almeida:IEEE2001}, where hysteresis is assumed to exist in different domains of the system under investigation. Here the hysteresis widths and amplitudes, and distribution of domains are fitting parameters of the model\cite{Almeida:APL2004,Almeida:IEEE2001}. To the best of our knowledge, there is a scarcity of physical models of the hysteresis and a discussion of its microscopic origin, especially for first order phase transitions. 
	
	
	A powerful tool for studying the hysteresis in various system is the first-order reversal curve (FORC) analysis \cite{Katzgraber,pike:6660}. FORC diagrams are based on the procedure first-described by Mayergoyz \cite{Mayergoyz}. They are used extensively in the study of magnetic systems, providing information regarding different phases\cite{Davies:PRB2005},
interactions\cite{Pike:JAP1999} and magnetic state of the system. examples include the identification of a transition from single domain to a vortex state in nanoscale magnets \cite{Dumas:PRB2008,Dumas:APL2007}, and one dimensional bubble states which exhibit large coercive fields in Co/Pt multilayers \cite{Davies:PRB2004}. The FORC method can be extended to any system where the response to an external parameter shows hysteretic behavior\cite{enachescu:054413}. This has been utilized in the study of ferroelectric materials\cite{Stancu:APL2003}, spin-crossover solids\cite{Tanasa:PRB2005} and spin glasses\cite{Katzgraber:PRL2002}.
	
	In this paper we report on our FORC measurements of the temperature driven MIT in VO$_2$ thin films. We prove the FORC analysis to be a useful tool in quantifying the hysteresis of different samples, which is important in any attempt to correlate the hysteresis parameters with various sample characteristics. Moreover, we find that even at low temperatures (down to 321K), where the VO$_2$ films seem to have no hysteresis, there is an irreversible signal in the FORC analysis. This feature is attributed to persistence of metallic domains down to these low temperatures, and indicates that there are interactions between the metallic domains and the surrounding films during the phase  transition. Assuming that such interactions should exist throughout the phase separated transition of the VO$_2$ film\cite{Qazilbash:science2007,sharoni:026404}, we propose a phenomenological model of the hysteresis in VO$_2$ samples. We take the interactions between the different VO$_2$ domains as the driving force behind the opening of the hysteresis, which is fundamentally different from the various forms of the Preisach model, since it does not assume the existence of hysteresis in every domain. This model enables us to reconstruct the entire FORC diagram from one full heating branch and two fitting parameters.

In Sec.~\ref{exp} the experimental setup and procedures are described in detail. In Sec.~\ref{Res} we show the experimental data, along with a physical interpretation. In Sec.~\ref{theory} we describe our theoretical model of thermal hysteresis, and Sec.~\ref{summary} is devoted to a summary of our findings.

\section{Experimental}\label{exp}
\subsection{Sample preparation}\label{exp1}

Vanadium oxide thin films were prepared by reactive RF magnetron sputtering of a
vanadium target (1.5" diameter, 99.8\%) on r-cut ($10\overline{1}2$)
sapphire substrates.  The samples were prepared in a high-vacuum deposition system with a
base pressure of 5$\times$10$^{-8}$ Torr.  A mixture of ultra-high purity (UHP) Argon and
UHP oxygen gasses were used for sputtering.  The total pressure during deposition was 3$\times$10$^{-3}$ Torr, and the oxygen partial pressure was optimized to 1.5$\times$10$^{-4}$ Torr (5\% of the total pressure).  The substrate temperature during deposition was 500 $^{\text{o}}$C; while the effective RF-magnetron power was kept at 300W.  These conditions yielded a deposition rate of 0.37 \AA/s.  The samples were cooled at a rate of 13 $^{\text{o}}$C/min in the same Ar/O$_2$ flow of the deposition.  Films were characterized and verified to be single phase VO$_2$ by X-ray diffraction using CuK$_{\alpha}$  radiation and by Energy-dispersive X-ray spectroscopy. Surface morphology was measured by Atomic force Microscopy (AFM) and Scanning Electron Microscope (SEM).

\begin{figure}
    \centering
        \includegraphics[width=0.35\textwidth]{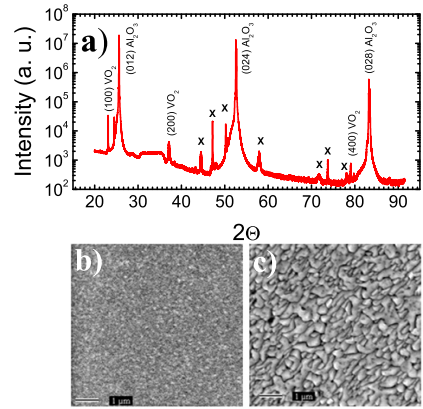}
    \caption{a) Room temperature, X-ray diffraction pattern of a 100nm VO$_2$ sample on a r-cut ($10\overline{1}2$) sapphire substrate.  ($\ell$00) planes of the M1 phase are observed.  Substrate peaks are marked along with instrumental background peaks (x-marked peaks).  b) SEM micrograph (7 $\mu m$ side) of the 100nm sample (small grain $~40nm$) and c) the 150nm sample (larger grain $~130nm$).}
    \label{fig:XRD}
\end{figure}

In this paper, we report on the properties of two films with different thicknesses, 150nm and 100nm.  The room temperature, X-ray diffraction pattern for the 100nm film is displayed in Fig.\ref{fig:XRD}(a).  Diffraction patterns for both films show similar features.  A large signal is measured for all relevant orders of the ($\ell$00) peak, confirming excellent monoclinic structure of VO$_2$, found for r-cut sapphire substrates \cite{Garry:TSF2004}.  No other vanadium oxide peaks were observed.
The grain size distribution extracted from the SEM images (see (Fig.\ref{fig:XRD}(b)) and (c) was $\sim$40nm for the 100nm film, and $\sim$130nm for the 150nm-thick film.  Variation in grain size is due to the difference in sample thickness and probably also to intricate differences in sample preparation, such as the exact deposition temperature \cite{Garry:TSF2004,Brassard:APL2005,Youn:JVST:2004}.

\subsection{Experimental Setup}\label{exp2}

Thermal hysteresis properties of thin VO$_{2}$ films were studied from analysis of resistance vs. temperature measurements by using a FORC procedure (see below).  The resistance characteristics were acquired in a home-made test bed by standard 4-probe measurements using a constant current source and a voltmeter at a sampling rate of 10 Hz.  A temperature controller was used to sweep the temperature at a constant ramp rate of 3K/min.  In this experiment, it is crucial to avoid temperature oscillations during sweeps, and especially during changes in the sweep direction.  We used two Platinum temperature sensors to avoid any such effect: One sensor was positioned close to a resistive heater, and provided feedback to the control loop.  The close vicinity to the heater reduces temperature oscillations in the sample which is farther away.  The second sensor was in thermal contact with the sample.  This sensor provided accurate measurements of temperature during the thermal cycles, along with the exact reversal temperature of each reversal curve, see below.

\subsection{FORC procedure}\label{exp3}
\begin{figure}
    \centering
        \includegraphics[width=0.45\textwidth]{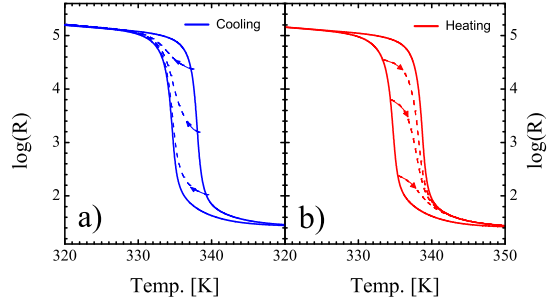}
    \caption{A set of first-order reversal curves.  a) Heating FORC.  b) Cooling FORC.  Solid lines are major loops, between fully insulating and fully metallic states.  Doted lines show reversal curves $\log R\left(T_R,T\right)$ for three different $T_R$.}
    \label{fig:fig1}
\end{figure}

A set of reversal curves for each sample was obtained as follows: starting at a temperature of 380 K (fully metallic state), the temperature was ramped down to a specific reversal temperature, $T_{R}$.  When $T_{R}$ is reached, the sample is heated back to the fully metallic state.  The path followed by the resistance, starting at $T_{R}$ and ending at 380K ($R(T_{R},T)$), is called a first-order reversal curve.  $T_{R}$ was varied between 370K and 260K in steps of $\sim 2$K around the saturated regions; while in the middle of the MIT, we used $T_{R}$ steps of $\sim 0.05$K.  Approximately 200 FORCs were obtained for each measurement set.
FORC measurements obtained with this procedure are called Heating FORCs, because temperature always increases for each $T_{R}$, see Fig.\ref{fig:fig1}(a).  Similarly, we can define Cooling FORCs (Fig.\ref{fig:fig1}(b)) by beginning with the fully insulating state, then heating up to various reversal temperatures, and finally obtaining a FORC for each $T_{R}$ by cooling back to the insulating state.  Herein, we show results for heating FORCs.  To rule out any spurious effects, due to sample degradation with measurements, we measured FORC sets with both increasing and decreasing $T_{R}$ between sequential curves.  No difference was found between the two procedures.
For every set of data we calculated the FORC distribution, defined as the mixed second-order derivative\cite{pike:6668},
\begin{equation}
    \rho\left(T_R,T\right)=-\frac{1}{2}\frac{\partial ^{2}r(T_R,T)}{\partial T_R \partial T},
\end{equation}
where  $r\left(T_{R},T\right)\equiv\log{\left(R\left(T_{R},T\right)\right)}$.  By calculating the FORC distribution for the log of the resistance there is an increase in sensitivity to features over the entire range of resistance, which is almost four decades  \cite{Almeida:IEEE2001}.
The mixed second-order derivative eliminates the parts in the data set where resistance is constant under a change in T or in $T_{R}$.  Thus, any non-zero value in the distribution corresponds to irreversible parts in the hysteresis loop.

\section{Results}\label{Res}
\begin{figure*}
\centering
\includegraphics[width=0.7\textwidth]{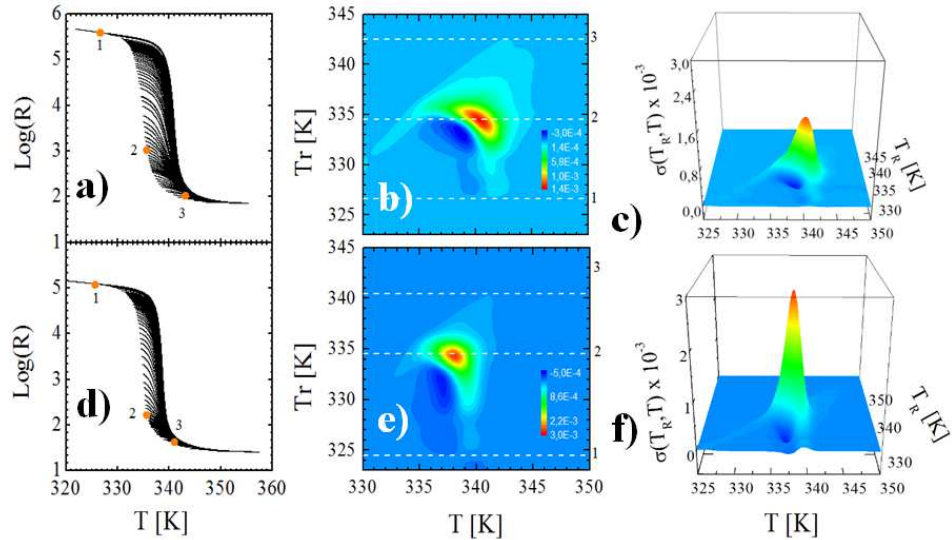}
\caption{Complete set of First-order reversal curves, the corresponding contour plot distributions, and 3D view for the VO$_2$ films on r-cut sapphire substrate.  Top row corresponds to the 100nm-thick sample (40nm grain size) and the bottom row to 150nm-tick (130nm grain size).}
\label{fig:FORC}
\end{figure*}

The heating FORC data sets measured for our VO$_2$ thin films are presented in Fig.\ref{fig:FORC}(a) for the thin film sample and Fig.\ref{fig:FORC}(d) for the thick film one.  The major hysteresis loops define boundaries of the reversal curves.  The corresponding FORC distributions are depicted in the contour plot in Fig.\ref{fig:FORC}(b) and \ref{fig:FORC}(e) for the thin and thick samples, accordingly.  Figures \ref{fig:FORC}(c) and \ref{fig:FORC}(f) are 3D representations of these distributions, with identical z-scales, enabling direct comparison of magnitudes in the FORC distributions between the two samples. 

While the general features for both samples are similar, the FORC distribution contours enable us to easily compare various aspects of the two samples measured, and to also find some surprising features in the reversal curves, which are hard to identify otherwise.  Three dotted lines are marked in Fig.\ref{fig:FORC}(b) and (e) to illustrate different reversal stages in the samples.  The reversal temperatures of these curves are marked in Figures \ref{fig:FORC}(a) and \ref{fig:FORC}(d).  These points represent the starting point of a specific curve.

The samples are insulating and fully reversible for temperatures below line 1 (in figures \ref{fig:FORC}(b)-100nm sample and \ref{fig:FORC}(e)-150nm sample); hence, there is no reversal signal.  This line corresponds to 327K in the 100nm sample and 324K in the 150nm one. Above line 1, a non-zero part evolves as $T_{R}$ is increased, appearing as a tail composed of a positive-negative part, around $T=340\text{K}$.
    
Interestingly, at point 1 in the FORC curves (presented in figures \ref{fig:FORC}(a) and (d)), it seems there is no hysteresis for this reversal temperature.  This feature is clear in the 150nm sample (Fig.\ref{fig:FORC}(d)) where the hysteresis opens only 4 degrees higher, around 328K.  But this point corresponds to the $T_{R}$, where the tail starts, i.e., where there is an irreversible feature in the FORC distribution.  This feature and its meaning will be discussed in detail further below.

\begin{figure}
    \centering
        \includegraphics[width=0.35\textwidth]{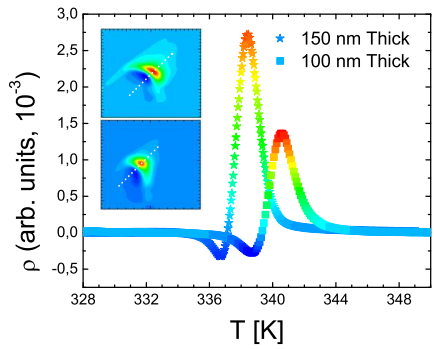}
    \caption{Profile lines parallel to $\text{T}=\text{T}_{\text{R}}$ axis, passing through the $\rho_{MAX}$ marked in the FORC distribution (inset) for the 100nm sample (top) and the 150nm sample (bottom).  Main panel - 100 nm sample (square-symbol) and 150 nm (star-symbol). Curves with same color map as the FORC plot. }
    \label{fig:lines}
\end{figure}

As $T_R$ increases, the tail joins a dip-peak structure which is the main part of the hysteresis, and includes most of the irreversible structure.  The peak maximum is marked by line 2, around $T_{R}=334\text{K}$ for both samples.  There are many quantitative properties of the hysteresis and reversal process that one can extract from the FORC procedure.  It is clear that the peak for the 100nm sample is wider and shifted
to higher temperatures versus the 150nm sample.  Figure \ref{fig:lines} illustrates this point, showing a cross-section that passes through the FORC peak and is parallel to $T=T_{R}$ (see inset of Fig.\ref{fig:lines}).  The square shaped plot is for the 100nm sample and the star shaped one is for the 150nm sample.  We find the maximum peak for the thin sample occurring at T$=$340.5K with magnitude of $\rho_{\text{max}}\approx1.4\times10^{-3}$ and full width half maximum (FWHM) of 1.60.  For the 150nm sample the transition is shifted to lower temperatures with the maximum at $T=334.6\text{K}$, a larger maximum of $\rho_{\text{max}}\approx2.74\times10^{-3}$
and narrow distribution, FWHM of $1.30$.  The negative part of the FORC distribution in both samples has similar features regarding minimum and width.

Above line 3, in Fig.\ref{fig:FORC}(b) and \ref{fig:FORC}(d), the FORC distribution is again zero, indicating that all R-T curves are completely reversible above this temperature and the sample is fully metallic.  These temperatures correspond to $343\text{K}$ and $340\text{K}$ for the 100nm and 150nm samples, respectively.

Thus, the FORC analysis presented above provides much quantitative information of the samples studied.  For example, Ref. \onlinecite{BrassardAPL2005} reports that the hysteresis of a VO$_2$ thin film sharpens as the grains become larger, which agrees with our findings.  By performing FORC measurements and analysis, one can quantitatively compare additional features of the hysteresis with various film characteristics.  These include the center of the hysteresis at T and $T_R$, the FWHM, and temperature limits of the reversible features. Such quantifying capabilities are valuable tools for analysis and understanding of the relationship between these properties.

More importantly, the FORC analysis enables us to observe features in the hysteresis, which are difficult to detect otherwise.  As stated earlier, in both films there is a tail in the FORC distribution around $T=340$K, extending to reversal temperatures lower than 325K.  This feature is enhanced in the thicker (150nm) VO$_2$ sample.  Upon examining the major loops of the hysteresis curve in this sample, hysteresis begins to open between the major heating and cooling curves at $\sim 329$K.  But, for reversal temperatures between 325K and 329K there is a non-zero value in the reversal curves around $T=340\text{K}$, leading to a non-zero value in the FORC distribution; and only for $T_R < 325K$ the FORC distribution depicts a fully reversible region.

This feature is similar to the one reported by Davis \textit{et al.} \cite{DaviesPRB} for magnetic hysteresis in Co/Pt multilayers.  We follow a similar interpretation of our results.  VO$_2$ has been shown to have phase separation between metallic and insulating domains across the transition\cite{Qazilbash-science-2007,sharoni:026404}.  If there are metallic domains which can persist to low temperatures, then when we measure a reversal curve (from these low temperatures) their effect on the resistance is negligible since they are small.  But during the reversal curve, they may act as `seeds' for the transition, and - through interaction with neighboring sites - they can modify the path of the MIT compared with the path taken from a fully metallic state.  The change in path is manifested in the non-zero values and tail shape of the FORC distribution.

\begin{figure}
    \centering
        \includegraphics[width=0.35\textwidth]{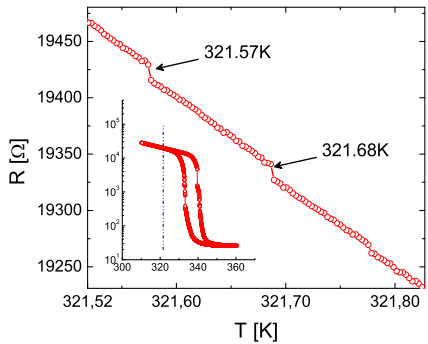}
    \caption{Resistance vs. Temperature for a $1\times6$~$\mu m^2$ VO$_2$ device.  Main panel shows a zoom around 321K of the major loop presented in the inset (dotted line indicates 321K). Large jumps, indicating domain switching, appear at 321.57K and 321.68K.}
    \label{fig:jumps}
\end{figure}

We find further reinforcement for this scenario by measuring the resistance vs. temperature characteristics of a cooling curve on a $1\times6$ $\mu m^2$ VO$_2$ junction with similar sample characteristics as the 150 nm sample, presented here.  We have previously shown that this measurement enables us to identify the transition of a single domain from metallic to insulating\cite{sharoni:026404}.  Figure \ref{fig:jumps} depicts such events as clear jumps in the resistance even at 321.6K.  This means that down to these temperatures, metallic domains exist in the sample.  The inset in Fig.\ref{fig:jumps} displays the full hysteresis loop of the device and the dotted line indicates the position of the jumps in the loop.  It is clear that metallic domains persist down to temperatures out of the hysteresis loop.

We wish to emphasize that the mere existence of metallic domains at lower temperatures should not change the shape of the FORC distribution unless these domains interact with the surrounding insulating domains, thus, affecting the course of the reversal curve.  These interactions possibly play an important role through the entire phase transition and could be the driving force in the opening of the hysteresis in the VO$_2$ MIT.  In the following section, we derive such a phenomenological theory and show that it can capture the main features we see in the FORC measurements.

\section{A Mean-Field Effective-Medium theory of hysteresis in VO$_2$}\label{theory}
Before proceeding, let us briefly describe the more common Preisach model \cite{Preisach:original,Almeida:IEEE2001}. The Preisach model, in its simplest form, is based on the \emph{assumption} that each grain has a transition temperature $T_c \pm \delta T_c$, where the plus and minus signs correspond to heating and cooling, respectively.  Averaging over $T_c$ and $\delta T_c$, the distributions of which serve as fitting parameters, one may obtain excellent fits between the model and the experimental data \cite{Almeida:APL2004}.

In the model presented here, we look for the microscopic origin of the hysteresis loop in the inter-grain interaction, rather than assuming that it is already present for the single grain.  The physical interpretation of our model is the following:  each grain has its own critical temperature, given by its parameters (e.g. size, oxygen doping etc.).  However, when the temperature exceeds the critical temperature of a specific grain, the grain still has to overcome an interaction-induced energy-barrier to change its phase.  The energy barrier depends on the concentration of same-type domains, which is the origin of the non-linearity required to generate hysteretic behavior.

The calculation method is as follows: first we calculate the concentration $p$ of metallic VO$_2$ domains as a function of temperature in the presence of inter-grain interaction for both directions of the temperature scanning (this corresponds to the mean-field part of the calculation). Once the concentration is calculated, it is used as input to calculate the resistance via the effective medium approximation (EMA)\cite{EffectiveMedium}.  The EMA relates the concentration of the different domains to the sample resistance, and has been successfully used to describe the MIT in VO$_2$ films \cite{rozen:081902}.  The resistance can be evaluated by using the EMA via the formula\cite{EffectiveMedium}
\begin{widetext}
\begin{equation}
    R = 2\left(\frac{1+p\left(1+z\right)}{R_{I}} +\frac{2+p\left(1+z\right)-z}{R_{M}} +
     \sqrt{\frac{4\left(z-2\right)R_{I}R_{M} + \left(\left(2+p\left(z-1\right)-z\right)R_{I} +
      \left(1+p\left(z-1\right)\right)R_{M}\right)^2}{R_{I}^2R_{M}^2}} \right)^{-1}\label{ema}
\end{equation}
\end{widetext}
where $R_I$, $R_M$ are the resistances of an insulating and metallic grain, respectively,  $z$ is the number of nearest neighbors, and $p$, as stated above is the concentration of metallic domains.

Let us start by calculating the concentration of metallic domains upon going from low to high temperatures- $p_{\text{up}}\left(T\right)$.  Consider first a single grain $\left(i\right)$ with its own critical temperature $T_{c}^{\left(i\right)}$. If there were no interactions, then the probability of finding this grain in a metallic state would simply be $p_{i}=\Theta\left(T-T_{c}^{\left(i\right)}\right)$, where $\Theta$ is the Heaviside step function. We now take into account the inter-grain interaction by assuming that the insulating domains around the $\left(i\right)$-th grain generate a potential barrier, which inhibits the transition.  This means that in order for the grain to turn metallic, it is not enough that $T>T_{c}^{\left(i\right)}$ , but there is also an energy barrier to overcome.  If the average interaction energy is $U$, then the effective energy barrier under a mean field approximation is $\Delta E^{i} = (1-p_{\text{up}})zU$, where $p_{\text{up}}$ is calculated by averaging over all the domains $\left\langle p_{\text{up}}^{i}\right\rangle$.  The probability of escape from a potential barrier is proportional to the Boltzmann factor, and so we have $p_{\text{up}}^{i}=\Theta\left(T-T_{c}^{\left(i\right)}\right)\text{exp}\left(-\frac{\left(1-p_{\text{up}} \right)zU}{k_{B}T}\right)$.  In order to average over all the domains, we assume that the critical temperatures $T_{c}^{\left(i\right)}$ of the different domains come from a Gaussian distribution averaged at the center of the transition with a typical width $\Delta T$.  With this assumption, we find the equation for $p_{\text{up}}$,
\begin{equation}
    p_{\text{up}}=\frac{1}{2}\left(1-\text{erf}\left(\frac{T-T_{c}}{\sqrt{2}\Delta T}\right)\right)
    \text{exp}\left(-\frac{\left(1-p_{\text{up}}\right)zU}{k_{B}T}\right)~~,
\end{equation}
which is solved numerically.  The cooling concentration $p_{down}$   may be calculated in the same way, now it is the fraction of metallic domains that generates the energy barrier.  In order to obtain reversal curves, we calculate $p_{down}\left(T\right)$ from an insulating state at high temperature to $T_R$, then use the concentration of metallic domains, and obtain $p_{\text{up}}\left(T,T_R\right)$ the concentration of metallic domains for a heating FORC with a reversal temperature $T_R$. Once $p_{\text{up}} \left(T,T_{\text{r}}\right)$ are obtained, the resistance is calculated via Eq.~\ref{ema}.

To fit the model to the experimental data, one needs the resistances of the metallic and insulating state.  However, these may be obtained experimentally from the low- and high-temperature data and are, thus, input parameters for the FORC analysis.  The only fitting parameters are the temperature distribution width $\Delta T$ and the interaction energy $U$ (assuming $z=4$ for simplicity).  Both of these parameters can be obtained only from the full hysteresis curve (i.e. solid line in Fig.~\ref{fig:fig1}).

\begin{figure}[t]
    \centering
        \includegraphics[width=0.34\textwidth]{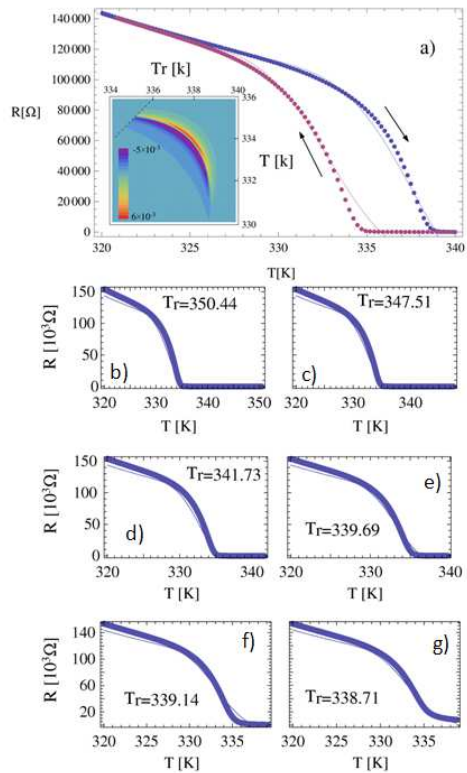}
    \caption{a) Resistance hysteresis curves obtained from the theoretical (solid lines) and experimental data (points).  The arrow direction indicates heating and cooling.  Inset: theoretical FORC curve, exhibiting similar features as the experimental one ("`moon-like"' shape, peak-dip structure).  b-g) Resistance vs. temperature for different reversal temperatures $T_{\text{R}}$. Solid lines are the theoretical curves and points are experimental data.  These curves were generated with no additional fitting parameters (see text).}
    \label{fig:theory}
\end{figure}

In Fig.~\ref{fig:theory}(a) we plot the theoretical curve R(T) for heating (blue line online) and cooling (red line online) measurements.  The points are the experimental data.  The parameters are $R_I(T)=36.6~\text{exp}\left(\frac{2647.5}{k_B T} \right) \Omega$, $R_M=25.7 \Omega$, $k_B \Delta T_c=2.26$K and $k_B U=79.85$K.  Already, this curve shows good agreement between the data and the theory (noting that resistance changes in four orders of magnitude).  In the inset of Fig.~\ref{fig:theory}(a) we plot the theoretical FORC curve, which  exhibits similarities to the experimental one, namely the "moon-like" shape and the peak-dip structure.

 In Fig.~\ref{fig:theory}(b-g) we plot the resistance curves for different values of $T_{\text{r}}$.  The theoretical curve is the solid lines, and the points are the experimental data.  We emphasize that these curves, which exhibit good fit between the theory and experiment, were obtained  without additional fitting parameters other then the two in Fig.~\ref{fig:theory}(a).

\section{Summary} \label{summary}
In summary, we presented FORC measurements and analysis of the hysteresis in the temperature driven metal insulator transition of vanadium oxide thin films.  We find that the FORC procedure, used extensively in characterizing magnetic systems is also a powerful tool for quantitative analysis of the MIT in VO$_2$.

The sensitivity of the FORC distribution to any irreversible feature enabled us to find that there are long-lasting metallic domains in the phase separated VO$_2$ film, existing down to temperatures where the sample appears fully insulating.  Moreover, we find a signature in the FORC analysis, indicating that these domains interact with the surrounding medium, modifying the path of the transition. 

In this spirit, we developed a mean field theory to capture the hysteresis.  Here, the driving force in charge of opening the hysteresis of the MIT is the interaction between nearest neighbor domains and the two fitting parameters having physical meanings.  This is enough to reconstruct the main experimental features. 

We hope this paper will encourage the use of the FORC method in other systems where appropriate, and that it will add to the theoretical discussion regarding the origin of hysteresis in general. 

\begin{acknowledgments}
We thank Randy K. Dumas and Kai Liu for fruitful discussions. This work was supported and funded by the US Department of Energy, AFOSR, and the Colombian agencies COLCIENCIAS and the Excellence Center for Novel Materials, CENM.
\end{acknowledgments}

\end{document}